%% file: paper.tex
\magnification 1200
\input my.macros
\title {Asymmetric Systematic Errors}
\docnum {MAN/HEP/03/02}
\date{3/6/2003}
\author {Roger Barlow}
\abstract {
\input abstract.tex}

\maketitle

\section Introduction

Although most errors on physics results are Gaussian, 
there are occasions where the Gaussian form no longer holds, and 
indeed when the distribution is not even symmetric. 

This can occur for statistical errors, when the one-$\sigma$ interval
is read off a log 
likelihood curve which is not well described by a parabola [1].
It can also arise
in evaluating systematic errors: if a `nuisance parameter' $a$
which affects the result $x$ has an uncertainty
described by a Gaussian distribution
with mean $\mu_a$ and standard deviation $\sigma_a$,
then
the uncertainty in $a$ produces an uncertainty in $x$
given 
to first order 
by the standard combination of errors formula:
$$\sigma_x^2 = \left( {dx \over da }\right)^2 \sigma_a^2.$$
The uncertainty in $a$ may be frequentist (for example, a 
Monte Carlo parameter determined by another experiment) or
Bayesian (for example, a Monte Carlo parameter set by 
judgement of theorists.)
Bayesian probabilities may be
admissable even in basically frequentist analyses if the effects are small [2].
The assumption
that $a$ has a Gaussian probability distribution may be questioned,
but that brings in further complications we do not wish to consider
here.

If the differential is not known analytically
a numerical evaluation can be done, most conveniently
by evaluation of $x(\mu_a+\sigma_a)$ and $x(\mu_a-\sigma_a)$.
See [3] for a discussion of the procedure and some issues that
may arise.

Both  $x(\mu_a+\sigma_a)-x(\mu_a)$ and $x(\mu_a)-x(\mu_a-\sigma_a)$
give estimates of the uncertainty $\sigma_x$. If they are 
different then this is 
a sign that the dependence is non-linear and the symmetric distribution in $a$
gives an asymmetric distribution in $x$.

The questions that can be asked are:

\bull How should asymmetric errors be combined?

\bull How should a $\chi^2$ be formed?

\bull How should a weighted mean be formed from results with asymmetric errors?

Current practice 
is to combine such errors separately, i.e. 
to add the $\sigma^+$ values together in quadrature, and then
do the same thing for the $\sigma^-$ values. This is not, to my knowledge,
documented anywhere and, as will be shown, is certainly wrong.

\section Models

The analysis gives 3 co-ordinate pairs: $(a-\sigma_a,x-\sigma_x^-)$,$(a,x)$
and $(a+\sigma_a,x+\sigma_x^+)$.
In practice there are errors on these points, and one might be well 
advised to assume a straight line dependence and take the error as
symmetric, however we will assume that this is not a case where this is 
appropriate. Again, faced with a real non-linear dependence 
one might well be advised to map out more than three points; we
will likewise assume that this is not done. We consider
cases where a non-linear effect is not small
enough to be ignored entirely, but not large enough to justify a 
long and intensive
investigation.
Such cases are common in practice.

For simplicity we transform $a$ to the 
variable $u$ described by a unit Gaussian, and work with $X(u)=x(u)-x(0)$.
For future convenience it is useful to define the mean $\sigma$, the difference $\alpha$, and the
asymmetry $A$:
$$\sigma = {\sigma^+ + \sigma^- \over 2}\qquad
\alpha = {\sigma^+ - \sigma^- \over 2}\qquad
A={\sigma^+ - \sigma^- \over \sigma^+ + \sigma^-}\eqno(1)$$

There are infinitely many non-linear
relationships between $a$ and $X$ that will go through these three points.
We consider two. 

\item {Model 1}: Two straight lines

Two straight lines are drawn, meeting at the central value
$$\eqalign{ X&=\sigma^+ u  \qquad u\geq 0 \cr
&=\sigma^- u  \qquad u \leq 0\cr}.\eqno(2)$$

\item {Model 2}: A quadratic function 

The parabola through the three points is

$$X=\sigma u + \alpha u^2=\sigma u + A\sigma u^2.\eqno(3)$$


%
%
%
%
%

%
These forms are shown in Figure 1 for a small asymmetry of 0.1, and 
a larger asymmetry of 0.4.
\vskip \baselineskip
\vskip 9 cm
\vbox{
\includegraphics{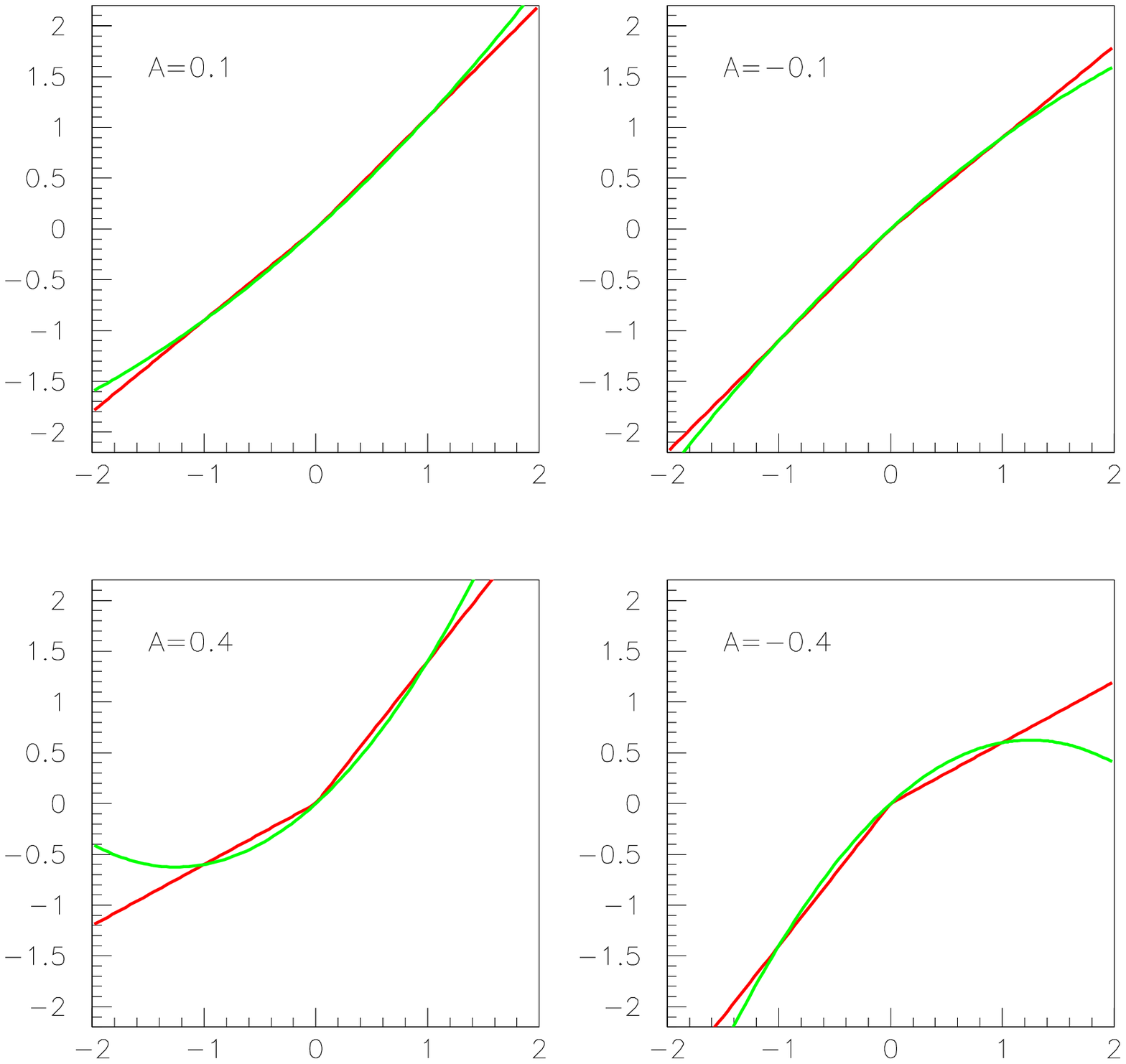}
\centerline{Figure 1: $X$ (vertically) 
against $u$ (horizontally)}
\vskip \baselineskip
}

Model 1 (two straight lines) is shown in red, and Model 2 ($x$ as a quadratic
function of $u$) in green.
Both go through the 3 specified points. The differences between them
within the range $-1\leq u \leq 1$ are not large; outside that range they
diverge considerably. 

We have no knowledge of whether either of them is
better than the other in a particular case. 
Model 1 has  kink at $u=0$ which is unphysical. Model 2 has a turning 
point, which may well be unrealistic (though it only gets into the
relevant region if $A$ is fairly large.)  
The practitioner may select one of the two - or some other model - on the
basis of their knowledge of the problem, or preference and experience. 
Working with asymmetric errors at all involves the assumption of 
some model for the non-linearity. The `correctness' of any model
may be
arguable, but once chosen it must be used consistently.

The distribution in $u$ is a unit Gaussian, $G(u)$, and the distribution 
in $X$ is obtained from
$P(X)={G(u) \over | dX/du |}$.   For Model 1 this gives a dimidated 
Gaussian -
two Gaussians with different standard deviation for $X>0$ and $X<0$
\footnote \dag{This is sometimes called a `bifurcated Gaussian', but this 
is inaccurate. 
`Bifurcated' means `split' in the sense of forked. 
`Dimidated' means `cut in half', with the subsidiary meaning
of `having one part much smaller than the other'[4].}.
For model 2 with small asymmetries the curve is a distorted Gaussian,
given by ${G(u) \over |\sigma + 2 \alpha u |}$ with 
$u={\sqrt{\sigma^2 + 4 \alpha X} - \sigma \over 2 \alpha}$. For larger asymmetries and/or
larger $|X|$ values, the second root also has to be considered. Examples
are shown in Figure 2.

\vskip \baselineskip
\vbox{
\vskip 9 cm
\includegraphics{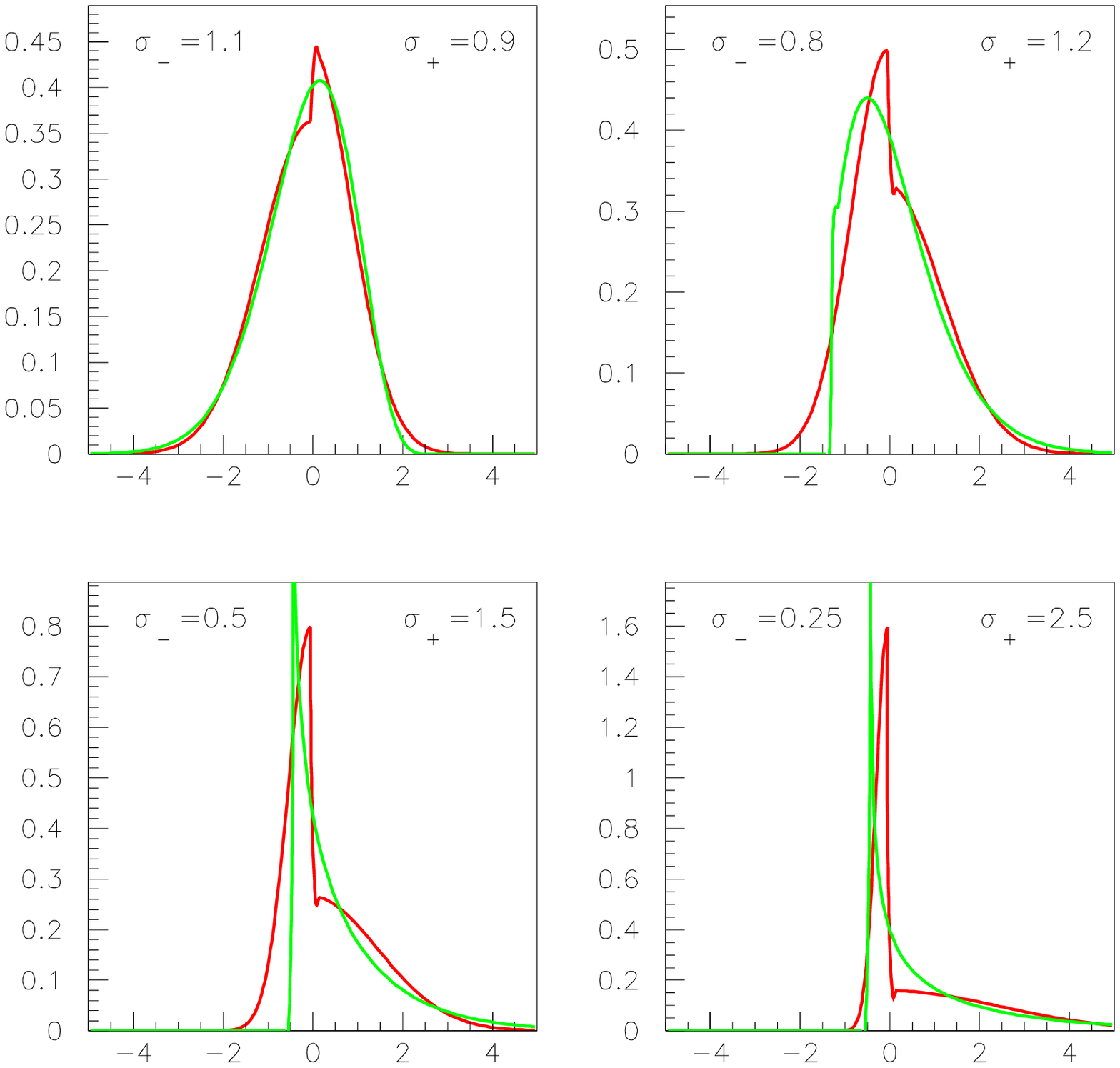}
\centerline{Figure 2: Examples of the distributions from combined asymmetric
errors. 
}
\vskip \baselineskip
}
It can be seen that the Model 1 dimidated Gaussian and Model 2 distorted Gaussian
are not dissimilar if the asymmetry is small, but are very different if the
asymmetry is large. Again, in a particular case
there is no unique reason for choosing one above the 
other in
the absence of further information.

\section Bias

If a nuisance parameter $u$ is distributed with a 
Gaussian probability distribution, and the quantity $X(u)$ is a 
nonlinear function of $u$, then the expectation $\langle X \rangle$
is not $X(\langle u \rangle )$.  

For model 1 one has $$<X> = 
\int_{-\infty}^0 \sigma^- u
{e^{-u^2 / 2 } \over \sqrt{2 \pi} } \,du
+\int_0^\infty \sigma^+  u
{e^{-u^2 / 2 } \over \sqrt{2 \pi} } \,du
 ={\sigma^+ - \sigma^- \over \sqrt{2 \pi}} \eqno(4)$$ 

For model 2 one has $$<X> = 
\int_{-\infty}^\infty 
\alpha
{u^2 e^{-u^2 / 2 } \over \sqrt{2 \pi} } \,du 
 ={\sigma^+ - \sigma^- \over 2 }  
 =\alpha\eqno(5)$$ 

Hence in these models, or others, 
if the result quoted is $X(0)$,
it is not the mean.  It is perhaps defensible
as a number to quote as the result as 
it is still the median - there is a 50\% chance that the true value is
below it and a 50\% chance that it is above.

\section Adding Errors

If a derived quantity $z$ contains parts from two quantities $x$ and 
$y$, so that $z=x+y$, the distribution in $z$ is given by the convolution:
$$f_z(z)=\int dx f_x(x) f_y(z-x)\eqno(6)$$

With Model 1
the function for $z\geq 0$ can be written:

$$f(z)=
\int_{-\infty}^0 dx f_{x-}(x) f_{y+}(z-x)
+\int_0^z dx f_{x+}(x) f_{y+}(z-x)
+\int_z^{\infty} dx f_{x+}(x) f_{y-}(z-x)$$

Inserting the appropriate Gaussian functions and using
$$
\sigma^2_{+}={\sigma_x^+}^2 + {\sigma_y^+}^2
\qquad
\sigma^2_{\pm}={\sigma_x^+}^2 + {\sigma_y^-}^2
\qquad
\sigma^2_{\mp}={\sigma_x^-}^2 + {\sigma_y^+}^2
\qquad
\sigma^2_{-}={\sigma_x^-}^2 + {\sigma_y^-}^2
, 
$$
this gives
$$\sqrt{2\pi}f(z)=
{1 \over \sigma_{\mp}} 
e^{-z^2\over 2\sigma^2_{\mp} }
g({-z \sigma_x^- \over \sigma_y^+ \sigma_{\mp}})
+{1 \over \sigma_{+}} 
e^{-z^2\over2\sigma^2_{+}}
\left(g({z \sigma_y^+ \over \sigma_x^+ \sigma_{+}})
-g({-z \sigma_x^+ \over \sigma_y^+ \sigma_{+}})\right)
+
{1 \over\sigma_{\pm}} 
e^{-z^2\over 2\sigma^2_{\pm}}
\overline {g}({z \sigma_y^- \over \sigma_x^+ \sigma_{\pm}})
$$
where $g(x)$ is the cumulative Gaussian, equivalent to ${1 \over 2}(1+erf(x))$,
and $\overline g(x)=1-g(x)$.

For $z\le 0$ the  limits are different and the
second region covers the case where $x$ and $y$ are both negative, giving
$$\sqrt{2 \pi} f(z)=
{1 \over \sigma_{\mp}} 
e^{-z^2\over2\sigma^2_{\mp} }
g({z \sigma_y^+ \over \sigma_x^- \sigma_{\mp}})
+{1 \over \sigma_{-}} 
e^{-z^2\over2\sigma^2_{-}}
\left(g({-z \sigma_x^- \over \sigma_y^- \sigma_{-}})
-g({z \sigma_y^- \over \sigma_x^- \sigma_{-}})\right)
+
{1 \over \sigma_{\pm}} 
e^{-z^2\over2\sigma^2_{\pm}}
\overline g({-z \sigma_x^+ \over \sigma_y^- \sigma_{\pm}})
$$

\vbox{
\vskip 9 cm
\includegraphics{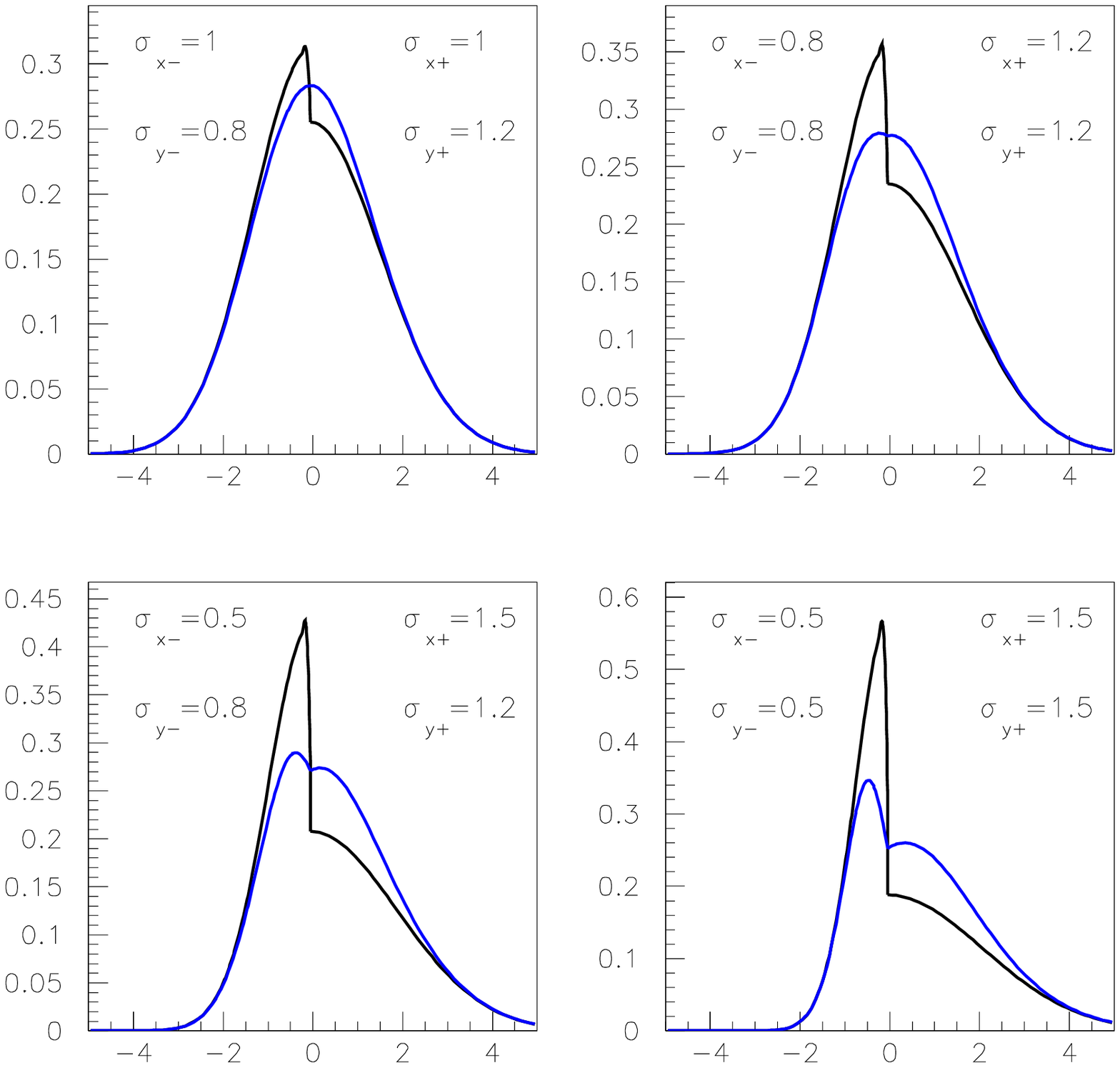}
\centerline{Figure 3: Examples of the distributions from combined asymmetric
errors. 
}
\vskip \baselineskip
}
Figure 3 shows the distributions from some typical cases. The
blue line shows the convolution, the black line 
is obtained by adding the positive and negative standard deviations separately
in quadrature (the `usual procedure').

The agreement is not good.
It is apparent that the skew of the distribution obtained from the convolution
is smaller than that obtained from the usual procedure. 
This is obvious: if two distributions with the same asymmetry are added
then the `usual procedure' will give a distribution with the same asymmetry.
This violates the Central Limit Theorem, which says that convoluting
identical distributions
must result in a combined distribution which is more Gaussian, and therefore
more symmetric,  than its components. This shows that the
`usual procedure' for adding asymmetric errors is inconsistent.  
Even though, as stated earlier, there is no guarantee that Model 1
or any 
model is correct, once a model has been adopted it should be handled in a
consistent fashion, and the `usual procedure' fails to do this.

\vfill \eject  

\section A consistent addition technique

If a distribution for $x$ is described by some 3 parameter function,
$f(x;x_0,\sigma^+,\sigma^-)$, which is a Gaussian transformed according to
 Model 1 or Model 2 or anything else, then `combination of errors' involves
a convolution of two such functions according to Equation 6.
This combined function is not necessarily a function of the same form. It is
a special property of the Gaussian that the convolution of two 
Gaussians gives a third.
Figure 3 is a demonstration of this. The convolution of two
dimidated Gaussians is not a dimidated Gaussian.

Although the form of the function is changed by a convolution, some things
are preserved. 
The semi-invariant cumulants of Thi\` ele (the coefficients of the power series expansion of the 
log of the Fourier Transform) add under convolution. The first two of these
are
the usual mean and variance. The third is the unnormalised skew:
$$\gamma = <x^3> - 3<x><x^2> + 2 <x>^3\eqno (7)$$

Within the context of any model, a rational approach to the combination
of
errors
is to find the mean, variance and skew: $\mu$, $V$ and $\gamma$, 
for each 
contributing function separately.  Adding these up gives the
mean variance and skew of the combined function. Working within the model
one then determines the values of
$\sigma_-, \sigma_+$, and $x_0$ 
that give this mean, variance and skew.

\subsection Model 1

For Model 1, for which
$\langle x^3 \rangle ={2 \over \sqrt{2 \pi}} (\sigma_+^3 - \sigma_-^3)$ we
have

$$\eqalign{\mu&=x_0+{1 \over \sqrt{2 \pi}} 
(\sigma^+ - \sigma^-)\cr
V &= {1 \over 2} ({\sigma^+}^2 + {\sigma^-}^2 ) - 
{1 \over 2 \pi} 
(\sigma^+ - \sigma^-)^2 = \sigma^2 + \alpha^2\left( 1-{2 \over  \pi}\right)\cr
\gamma&=
{1 \over \sqrt{2 \pi}} \left[
2
({\sigma^+}^3 - {\sigma^-}^3) 
-{3 \over 2}
({\sigma^+} - {\sigma^-}) 
({\sigma^+}^2 + {\sigma^-}^2) 
+{1 \over \pi}
({\sigma^+} - {\sigma^-}) ^3
\right]}
\eqno(8)
$$
So given a set of error contributions 
then the equations (8) give the cumulants $\mu$, $V$ and $\gamma$.
The first three cumulants  of the combined distribution  are given by
adding up the individual contributions.  Then one can find the set of
parameters
 $\sigma^-, \sigma^+, x_0$ 
which give these values by using Equations (8) in the other sense.

It is convenient to work with 
$\Delta$,
where $\Delta$ is the difference between the final $x_0$ and the
sum of the individual ones.
The parameter 
is needed because of the bias mentioned earlier.
Even though each contribution may have $x_0=0$, i.e. it describes a spread about the
quoted result, it has non-zero $\mu_i$ through the bias  effect
(c.f. Equation 4).
The $\sigma^+$ and $\sigma^-$ of the combined distribution, obtained from 
the total $V$ and $\gamma$, will in general not give the right $\mu$ unless 
a location shift $\Delta$ is added. {\it The value of the quoted result will shift.}

Recalling section 3, for a dimidated Gaussian one could defend quoting the  central value as it 
was the median, even though it was not the mean. 
The convoluted distribution not only has a non-zero mean, it also
(as can be seen in Figure 2) has non-zero median.
Consider two 
dimidated Gaussians with, say, 
$\sigma^+>\sigma^-$
, which are
convoluted. There is a 25\% chance that both will contribute a negative value,
a similar 25\% chance that both will be positive, 
and a 50\% chance of getting one
positive and one negative contribution  - which will probably be positive
overall 
(as  
$\sigma^+>\sigma^-$). 
So for  two combined 
distributions the zero value may lie as far away as 
the 25th percentile.
 
If you want to combine asymmetric errors 
then you have to accept that the quoted value will
shift.  To make this correction requires a real belief in the 
asymmetry of the error 
values. 
At this point the practitioner,
unless they are really sure that their errors really do have a
significant asymmetry,
may be
persuaded to revert to quoting symmetric errors.

Solving the Equations (8)  for 
 $\sigma^-, \sigma^+, x_0$ 
given 
$\mu$, $V$ and $\gamma$ 
has to be done numerically.  If we write
$D=\sigma^+ - \sigma^-$ and $ S = {\sigma^-}^2 + {\sigma^+}^2$
then the equations
$$\eqalign{S&=2 V + D^2/\pi\cr
D&={2 \over 3 S} \left(\sqrt{2 \pi} \gamma -D^3({1 \over \pi}-1)\right)}
\eqno(9)$$
can be solved by repeated substitution (starting with $D=0$). 
Then $\Delta$ is given by 
$$\Delta=\mu-{D \over \sqrt{2 \pi}}\eqno(10)$$
A program for this is available on {\tt http://www.slac.stanford.edu/$\sim$barlow}.
Some results are shown in Figure 4 and Table 1.
\vskip \baselineskip
\vbox{
\vskip 9 cm
\includegraphics{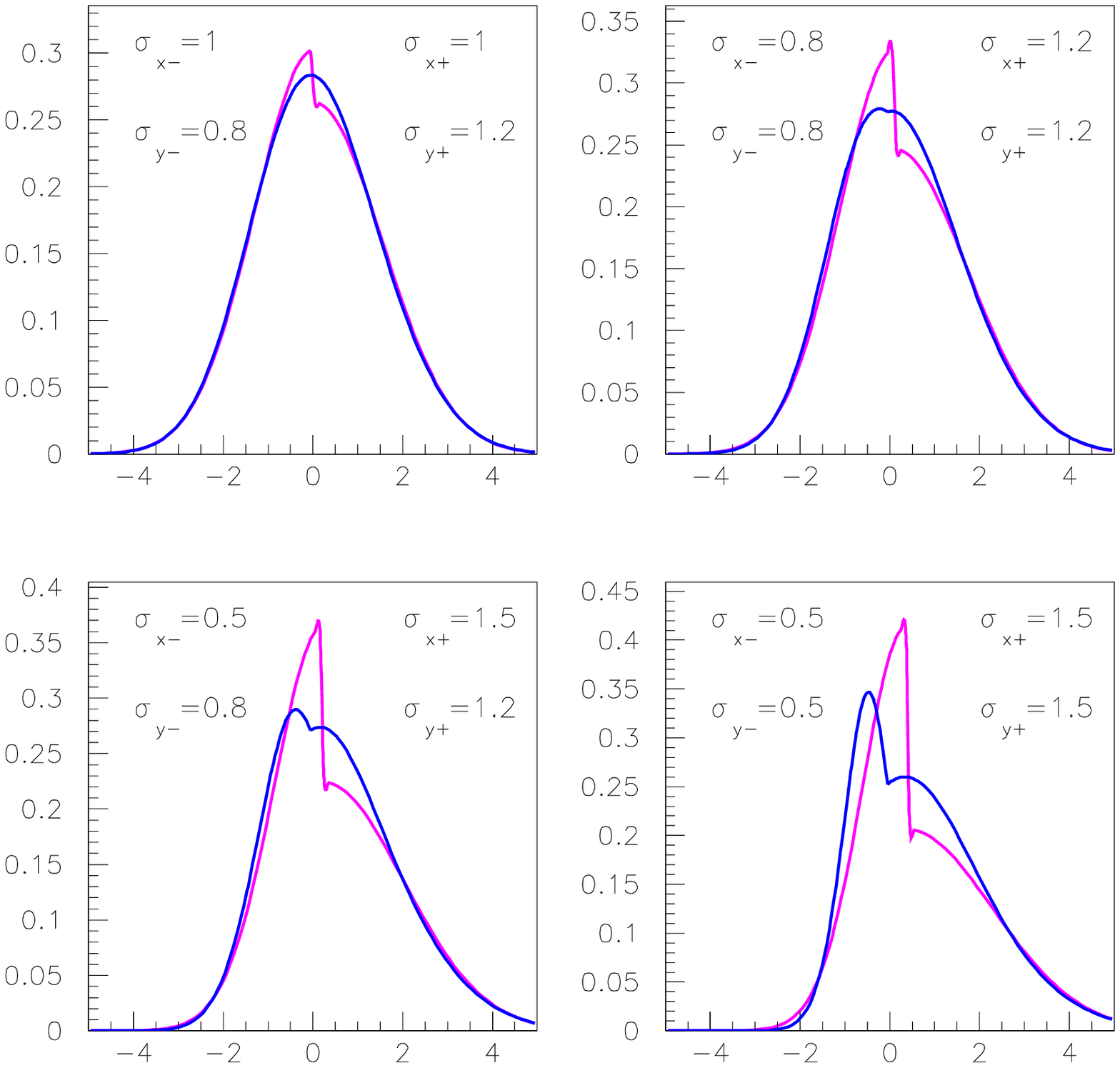}
\centerline{Figure 4: Examples of combined
errors with the correct first 3 cumulants using Model 1. 
}
\vskip \baselineskip
}

\centerline{
\vbox{\halign{
# & # & # & # & \qquad # & # & #   \cr
$\sigma_x^-$ &
$\sigma_x^+ $&
$\sigma_y^- $&
$\sigma_y^+ $&
$\sigma^{-} $&
$\sigma^{+} $&
$\Delta$\cr
1.0 & 1.0 & 0.8 & 1.2 & 1.32 & 1.52 & 0.08 \cr
0.8 & 1.2 & 0.8 & 1.2 & 1.22 & 1.61 & 0.16 \cr
0.5 & 1.5 & 0.8 & 1.2 & 1.09 & 1.78 & 0.28 \cr
0.5 & 1.5 & 0.5 & 1.5 & 0.97 & 1.93 & 0.41 \cr
}}}

\vskip \baselineskip
\centerline{Table 1: The values used in Figure 4}
\vskip \baselineskip

Comparing Figure 4 and Figure 3 (note that the blue curves are the
same in both figures; the consistent technique is shown in purple), 
it is apparent that the new technique 
does a very much better job than the old. It is not an exact match, but 
does an acceptable job given that there are only 3 adjustable parameters in the function.

\subsection Model 2

In terms of the difference $\alpha=(\sigma^+-\sigma^-)/2$ and the mean 
$\sigma=(\sigma^+ + \sigma^-)/2$
the moments are
$$<x>=\alpha \qquad <x^2>=\sigma^2 + 3 \alpha^2 \qquad  <x^3>=9 \alpha \sigma^2
 + 15\alpha^3$$
Giving 
$$\eqalign{\mu&=x_0+\alpha\cr
V&=\sigma^2 + 2\alpha^2\cr
\gamma&=6\sigma^2\alpha + 8 \alpha^3}\eqno(11)$$
As with Method 1, these are used to find the cumulants of each contributing distribution,
which are summed to give the three totals, and then Equation 11 is 
used again to find the
parameters of the distorted Gaussian with this mean, variance and skew.
There is only one equation to be solved numerically, again
by iteration
$$\alpha={\gamma  \over 6 V -4 \alpha^2}\eqno(12)$$
after which, $\sigma=\sqrt{V-2\alpha^2}$ and $\Delta = \mu-\alpha$.

Some results are shown in Figure 5 and Table 2. The true convolution
cannot be done analytically but can be done by a Monte Carlo calculation.
\vskip\baselineskip
\vbox{
\centerline{\vbox{\halign{
# & # & # & # & \qquad # & # & #   \cr
$\sigma_x^-$ &
$\sigma_x^+ $&
$\sigma_y^- $&
$\sigma_y^+ $&
$\sigma^- $&
$\sigma^+ $&
$\Delta$\cr
1.0 & 1.0 & 0.8 & 1.2 & 1.33 & 1.54 & 0.10 \cr
0.8 & 1.2 & 0.8 & 1.2 & 1.25 & 1.64 & 0.20 \cr
0.5 & 1.5 & 0.8 & 1.2 & 1.12 & 1.88 & 0.35 \cr
0.5 & 1.5 & 0.5 & 1.5 & 1.13 & 2.07 & 0.53 \cr
}}}
\vskip\baselineskip
Table 2: The values used for the curves with correct cumulants in Figure 5. 
\vskip\baselineskip
}

\vbox{
\vskip 9 cm
\includegraphics{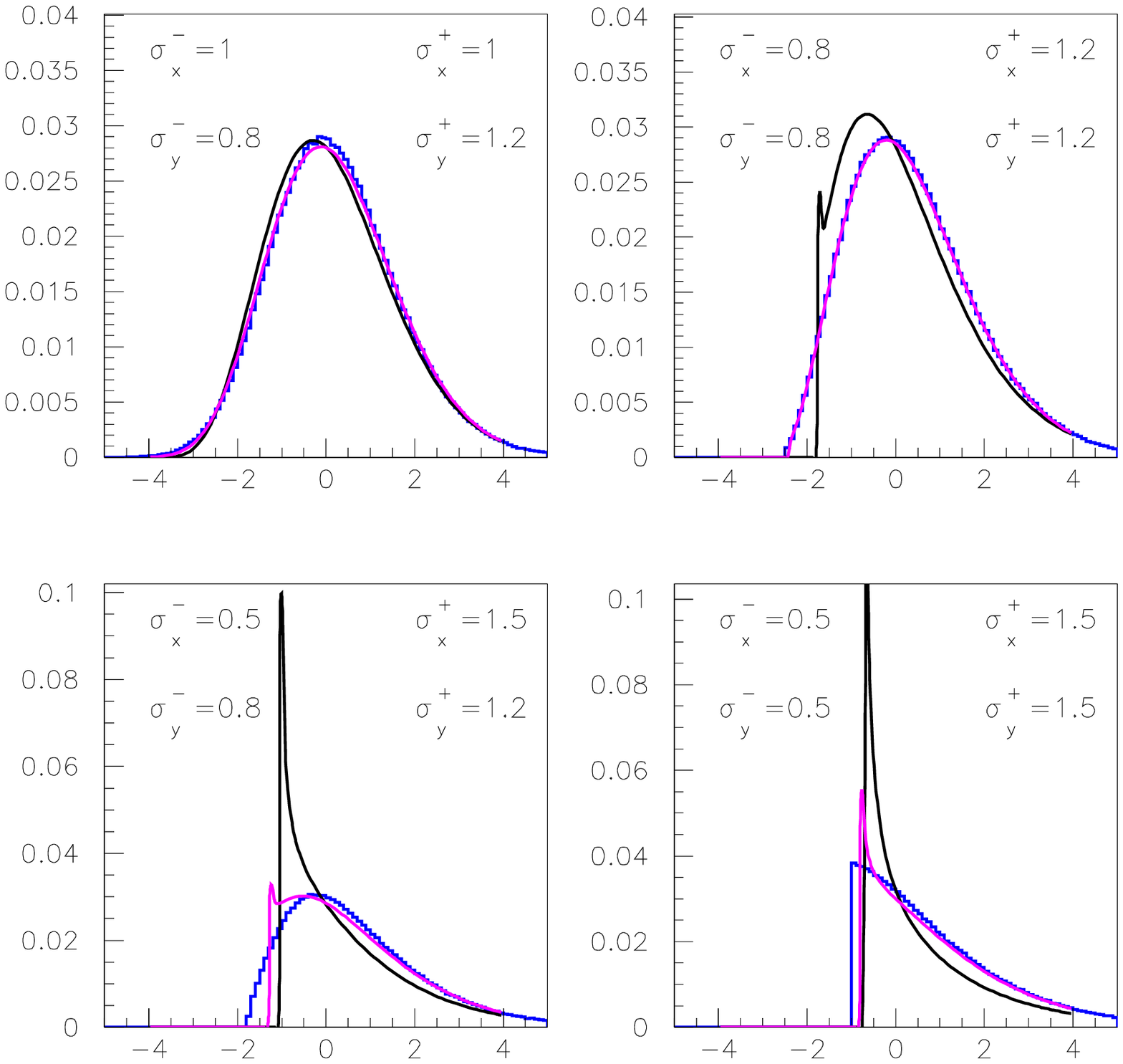}
\centerline{Figure 5: Examples of combined
errors using Model 2. 
}
}
\vskip\baselineskip

Again the true curves (blue) are not well reproduced by the `usual procedure' (black)
whereas the curves with the correct cumulants (purple) do a very
reasonable job. 
(The sharp behaviour at the lower edge of the curves is 
due to the minimum value
of $y$.)
 
The web program mentioned earlier will also do the calculations for Model 2.

\section Evaluating $\chi^2$

For Model 1 the $\chi^2$ contribution from a discrepancy $\delta$ is 
just $\delta^2/{\sigma^+}^2$ or $\delta^2/{\sigma^-}^2$ as appropriate.  
This is
manifestly inelegant, especially for minimisation procedures as the value goes through zero. 

For Model 2 one has
$$\delta=\sigma  u + A \sigma u^2$$.

This can be considered as a quadratic for $u$ with solution
$$u={\sqrt{1+4 {\delta \over \sigma}  A} -1 \over 2 A}$$
Squaring gives $u^2$, the $\chi^2$ contribution, as
$$u^2={2+4A{\delta \over \sigma} -2 (1+4A{\delta \over \sigma})^{1 \over 2}\over 4 A^2}$$
This is not really exact, in that it only takes one 
branch of the solution, the one approximating to
the straight line, 
and does not consider the extra possibility that the $\delta$ value could come
from an improbable $u$ value the other side of the turning point of the parabola.
Given this imperfection it makes sense to 
expand the square root as a Taylor series, which, neglecting correction
terms above 
the second power, leads to
$$\chi^2=({\delta \over \sigma})^2 \left(1-2A ({\delta \over \sigma})+
5 A^2 ({\delta \over \sigma})^2\right).\eqno(13).$$
The first order approximation  to this is 
$$\chi^2=({\delta \over \sigma})^2 (1-2A ({\delta \over \sigma})).\eqno(14)$$

This can be modified to a form forced to give $\chi^2=1$ for
deviations of $+ \sigma^+$ and $- \sigma^-$.  
$$\chi^2=\delta^2 ({{\sigma^+}^3+{\sigma^-}^3 \over {\sigma^+}^2 {\sigma^-}^2
({\sigma^+} + {\sigma^-})}) 
(1- \delta {{\sigma^+}^2- {\sigma^-}^2 \over {\sigma^+}^3 + {\sigma^-}^3})\eqno(15)$$

\vbox{
\vskip 8 cm
\includegraphics{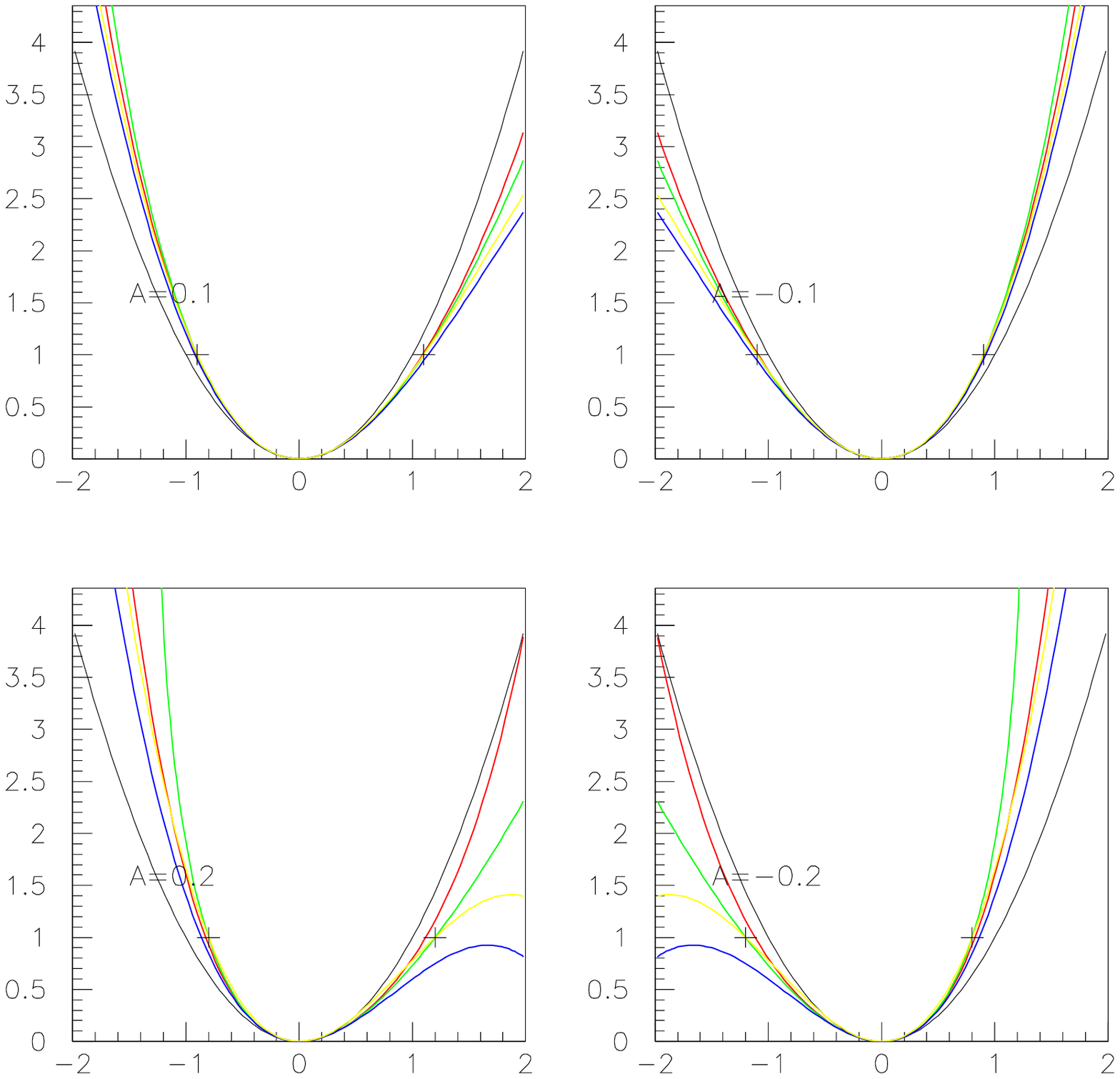}
\centerline{Figure 6: $\chi^2$ approximations }
}
\vskip\baselineskip

Figure 6 shows these forms.  The black line is the simplest $\chi^2=({\delta \over \sigma})^2$ form. 
The green is the full form involving the square root.
It goes to $+\infty$ for values beyond the turning point which in principle
can never happen.  The blue line is the third order form of Equation 14
and the red line is the higher order Equation 13.  
The yellow is Equation 15, the 
first order form constrained to go though unity at $+\sigma^+$ and $-\sigma^-$,
shown by the two crosses.
For a 10\% asymmetry all the approximations are pretty well equivalent
and a significantly better form than the simplest one. 
For a larger 20\% asymmetry the lower order forms show undesirable 
behaviour, turning over for
a moderate (2$\sigma$) deviation.

We therefore suggest that 
Equation 13 be used. 
The even power ensures that 
$\chi^2$ does not turn over but increases  at large deviations, which 
is desirable.
It does not go to infinity when $\delta$ approaches the turning point, 
which is probably a good feature.
A poor determination of the parameters of Equation 3 could give an unrealistic
minimum value which could be exceeded by an experimental value,
 and one would not want this to give an undefined 
$\chi^2$.

Higher order (5th, 6th...) terms do not significantly improve the agreement with
the full (green) curve.

\section Weighted means

Suppose a value $x$ has been
measured several times, $x_1,x_2...x_N$, each measurement having
its own $\sigma_i^+$ and $\sigma_i^-$.
For the usual symmetric errors the `best' estimate (i.e. unbiassed and with
smallest variance) is given by the weighted sum
$$\hat x ={ \sum w_i x_i \over \sum w_i}$$
with $w_i = 1/\sigma_i^2$. We wish to find the equivalent for asymmetric
errors. 

As noted in Section 3, when sampling from an asymmetric distribution the 
result is biassed
towards the high tail. The expectation value $\langle x \rangle$
is not the location parameter $x$. 
So for an an unbiassed estimator one has to take
$$\hat x= \sum w_i(x_i-b_i) / \sum w_i\eqno(16)$$
where
$$b={\sigma^+-\sigma^- \over \sqrt{2 \pi}} \quad \hbox{(Model 1)} \qquad 
b=\alpha \quad \hbox{(Model 2)} \eqno(17)$$
The variance of this is given by
$$V={\sum w_i^2 V_i \over \left( \sum w_i \right)^2}$$
where $V_i$ is the variance of the $i^{th}$  measurement about its mean.

Differentiating with respect to $w_i$ to find the minimum gives
$${2 w_i V_i \over \left( \sum w_j\right)^2} - {2 \sum w_j^2 V_j \over \left( \sum w_j \right)^3}=0\qquad \forall i$$ 
which is satisfied by $w_i = 1/V_i$.
This is the equivalent of the 
familiar weighting by $1/\sigma^2$. 
The weights are given by (see Equations 8 and 11)
$$V=\sigma^2 +(1-{2\over  \pi})\alpha^2 \quad \hbox{(Model 1)}\qquad V=\sigma^2 + 2 \alpha^2\quad \hbox{(Model 2)}\eqno(18)$$. 

Note that this is not the ML estimator - writing down the likelihood in terms
of the $\chi^2$ of Equation 13  and differentiating does not give to a nice form -
so in principle there may be better estimators, but they will not have the simple form of a weighted sum.

 \section Asymmetric statistical errors

When the estimated value and range are obtained using a maximum
likelihood estimate and the shape of the log likelihood is not
parabolic,  
the one standard deviation limits are taken
as the points at which the log likelihood falls by 0.5 from its peak [1].

The treatment of these errors will be given 
in a subsequent publication. Although treatment of asymmetric errors 
involves, for both systematic and statistical errors, the mapping of the 
actual distribution onto a Gaussian one, there is a considerable difference of
interpretation. It is, however, worth pointing out that if two separate
statistical effects are combined - say two backgrounds from different 
sources - then the combined background is the simple arithmetic sum of the two
with no shift to the central value. This is because, for these statistical errors,
the value quoted is the mean.   




\section Summary

The treatment of asymmetric systematic errors cannot be based on
secure foundations, and if they  cannot be avoided they need careful
handling.
The practitioner needs to choose a model for the dependence, which
could be one of the two proposed here.

In combining asymmetric errors, the traditional procedure of adding
positive and negative values separately in quadrature is unjustifiable.
Instead, values should be determined which, within the limitations of
the model, give the correct mean, variance, and skew.
A program is available to do this 
on {\tt http://www.slac.stanford.edu/$\sim$barlow}.

The $\chi^2$ contribution for a value with asymmetric errors can be
represented by
$$\chi^2=({\delta \over \sigma})^2 \left(1-2A ({\delta \over \sigma})+
5 A^2 ({\delta \over \sigma})^2\right)$$
where 
$$\sigma = {\sigma^+ + \sigma^- \over 2}\qquad
A={\sigma^+ - \sigma^- \over \sigma^+ + \sigma^-}$$

In forming a weighted sum 
one should use 
$$\hat x = \sum {(x_i-b_i)\over V_i} / \sum {1 \over V_i}.$$
where the bias $b$ and Variance $V$ are given 
by Equations 17 and 18 above.

\vskip \baselineskip

\leftline{\bf References}

[1] W. T. Eadie et al, ``Statistical Methods in Experimental Physics'', 
North Holland, 1971

[2] R. D. Cousins and V. L. Highland, \NIM 320 331 (1992)

[3] R. J. Barlow ``Systematic Errors: Facts and Fictions" in Proc. Durham 
conference on Advanced Statistical Techniques in Particle Physics, 
M. R. WHalley and L. Lyons (Eds). IPPP/02/39. 2002

[4] 
The Shorter Oxford 
English Dictionary, Vol I (A-M) p 190 and p 551 of the 3rd edition (1977). 

\bye

%% file: abstract.tex
Asymmetric systematic errors arise 
when there is a non-linear dependence of a 
result on a nuisance parameter.
Their combination
is traditionally done by adding 
positive and negative deviations separately in quadrature. There is no
sound justification for this, and it is shown that indeed it is 
sometimes clearly inappropriate. Consistent techniques are
given for this combination of errors, and also  for
evaluating  
$\chi^2$, and for forming
weighted sums.